\documentclass[aps,prstab,groupedaddress,showpacs,showkeys,twocolumn,floatfix]{revtex4-1}

\usepackage{graphicx} \usepackage{url} \usepackage{amsmath}
\usepackage{color}
\usepackage{psfrag}

\begin{document}

\title{Coupled beam motion in a storage ring with crab cavities}

\author{Xiaobiao Huang }
\email[]{xiahuang@slac.stanford.edu} 
\affiliation{SLAC National Accelerator Laboratory, Menlo Park, CA 94025}

\date{\today}

\begin{abstract}
We studied the coupled beam motion in a storage ring between the transverse and longitudinal 
directions introduced by crab cavities. 
Analytic form of the linear decoupling transformation is derived. 
The equilibrium bunch distribution in an electron storage ring with a crab cavity 
is given, including contribution to the eigen-emittance induced by the crab cavity.
Application to the short pulse generation scheme using crab cavities~\cite{Zholents2015111} 
is considered. 
\end{abstract}
 

\maketitle

\section{\label{intro}Introduction}
Crab cavities (also known as transverse deflecting cavities) have found major applications in 
storage rings. In colliders, they are used to rotate the colliding bunches at the collision point to 
create head-on collisions while the trajectories of the two beams cross at an angle (crab crossing)~\cite{Oide.PRA.40.315}. 
In electron storage rings, it has been proposed to use crab cavities to tilt a long bunch in the 
$y$-$z$ plane in order to select a short X-ray pulse from the radiation generated by the beam with 
a vertical slit~\cite{Zholents1999385, Zholents2015111}. 

A crab cavity gives the beam a time-dependent transverse kick. 
The kick is typically in the horizontal plane for the crab crossing application 
and in the vertical plane for the short pulse application. 
By virtue of the Panofsky-Wenzel theorem, the crab cavity also gives the beam 
a longitudinal kick that is dependent on the transverse offset. 
Naturally the crab cavity couples the transverse direction to the longitudinal direction. The nature of the linear coupled 
motion between the $y$-$z$ or $x$-$z$ directions is the same as the linear $x$-$y$ coupling introduced 
by a skew quadrupole. Therefore it can be likewise studied. 

In some earlier work, the effects of the crab cavity on the beam are described as 
creating a $z$-dependent closed orbit~\cite{Yipeng.PRSTAB_2009,Zholents2015111}. 
Although it can lead to useful results, such a physics picture may be incorrect. 
Because a particle with a $z$-offset will undergo synchrotron motion, it does not 
stay on the orbit. 

In this paper we study the coupled motion due to crab cavities in a storage ring through the 
transfer matrix. A matrix perturbation method is applied to find the linear transformation 
that block diagonalizes the one-turn transfer matrix. Analytic formulas for the 
decoupling linear transformation are derived. 
By applying the  matrix perturbation technique to the Ohmi envelope equation~\cite{Ohmi.PRE.49.751} 
and considering the quantum diffusion of the beam with a tilted distribution on the $y$-$z$ directions, 
the equilibrium phase space distribution in an electron storage ring with crab cavities is also 
obtained. 
These results are applied to the short pulse generation scheme~\cite{Zholents2015111}. 
The short pulse performance is calculated and its functional dependence on crab cavity 
and lattice parameters  is revealed . It is shown that 
the vertical eigen-emittance due to the crab cavity induced tilt in bending magnets is a dominant 
factor that limits the achievable minimum pulse duration. 
Numeric example and simulation results for the short pulse generation application for SPEAR3 
are presented. 

In section~\ref{secyz} we study the decoupling transformation for a vertical crab cavity. 
In section~\ref{secYZOhmi} we first show the connection between the second order moment matrix 
of the original and decoupled coordinates. Then we derive the changes to the equilibrium distribution 
due to the crab cavity and calculate the short pulse performance for the short pulse generation scheme.  
A brief description of the procedure for decoupling the motion by a 
horizontal crab cavity is given in section~\ref{secXcrab}.
Numeric examples and particle tracking are shown in section~\ref{secSimul}. The conclusion is 
given in section~\ref{secConclu}.

\section{\label{secyz} Linear coupling by a vertical crab cavity}
For a conventional crab cavity working at the TM110 mode, assuming the transverse deflection is on the vertical 
direction, the E-M fields are given by 
\begin{eqnarray}
E_z = \mathcal{E}_0 k y \cos \omega t, \quad cB_x = \mathcal{E}_0 \sin \omega t,
\end{eqnarray}
where $\mathcal{E}_0/c$ gives the amplitude of the magnetic induction, $k=\omega/c$ the angular wave number, $\omega=2\pi f$ the 
angular frequency, and $c$ the speed of light. In crab cavity applications, the beam 
arrives at around $t=0$, where kick-to-time slope is the maximum.  
Correspondingly the kicks to the beam in linearized form are
\begin{eqnarray}
\Delta y'=\frac{eV}{E} k z, \quad \Delta \delta = \frac{eV}{E} k y,
\end{eqnarray}
where $\delta$ is the momentum deviation of the particle, 
$E$ is the beam energy and $V=\int_{\rm gap} \mathcal{E}_0 \sin (\omega t) cdt$ is the deflecting voltage. 
The linear motion through a crab cavity can be expressed via a transfer matrix 
of the coordinates ${\bf X}=(x,x',y,y',z,\delta)^T$.
For a thin vertical crab cavity,  the transfer matrix is given by 
\begin{eqnarray}\label{eqTcgen}
{\bf T}_{c} = 
\left( \begin{array}{ccc} {\bf I} & {\bf 0} & {\bf 0}  \\ 
{\bf 0} &  {\bf I} & \epsilon {\bf W} \\
{\bf 0} & \epsilon {\bf W}  &  {\bf I}  \end{array}\right),
\end{eqnarray}
where ${\bf I}$ in this paper is the identity matrix of the appropriate size, $2\times2$ in this case,  
$\epsilon = \frac{eVk}{E} $, and 
\begin{eqnarray}
{\bf W} = \left( \begin{array}{cc} 0 & 0 \\
1 & 0\\
\end{array}\right).
\end{eqnarray}

When a crab cavity is inserted into the ring lattice, the one-turn transfer matrix will be changed. 
Assuming the crab cavity is located at point 2, the one-turn transfer matrix at point 1 is
\begin{eqnarray}\label{eqT1gen}
{\bf T}_1 &= & {\bf T}_{12}{\bf T}_{c}{\bf T}_{21},
\end{eqnarray}
where 
${\bf T}_{21}$ is the transfer matrix from point 1 to 2 and ${\bf T}_{12}$ the transfer matrix from point 2 to 1.  
Using notations as defined in Ref.~\cite{Huang.MatrixXZ.PRSTAB}, the transfer matrix ${\bf T}_{21}$ can be written
\begin{eqnarray}
{\bf T}_{21} = 
\left( \begin{array}{ccc} {\bf M}_{x,21} & {\bf 0} & {\bf E}_{21}  \\ 
{\bf 0} &  {\bf M}_{y,21} & {\bf 0} \\
{\bf F}_{21} & {\bf 0}  &  {\bf L}_{21}  \end{array}\right),
\end{eqnarray}
where each element is a $2\times2$ matrix block. Transfer matrix ${\bf T}_{12}$ can be likewise expressed. 
Working out Eq. (\ref{eqT1gen}), we found
\begin{eqnarray}
{\bf T}_1 =  {\bf T}_{1}^{(0)} + \epsilon \tilde{\bf T}_1, 
\end{eqnarray}
with
\begin{eqnarray}
\tilde{\bf T}_1 =
\left( \begin{array}{ccc} {\bf 0} & {\bf E}_{12}{\bf W}{\bf M}_{y,21} & {\bf 0}   \\ 
{\bf M}_{y,12}{\bf W}{\bf F}_{21} & {\bf 0} &  {\bf M}_{y,12}{\bf W}{\bf L}_{21} \\
{\bf 0} &   {\bf L}_{12}{\bf W}{\bf M}_{y,21}  &{\bf 0}  \end{array}\right),
\end{eqnarray}
where ${\bf T}_{1}^{(0)}$ is the one-turn matrix at point 1 without the crab cavity, 
\begin{eqnarray}\label{eqT10}
{\bf T}_{1}^{(0)} &=& {\bf T}_{12}{\bf T}_{21} = 
\begin{pmatrix} {\bf M}_{x} & {\bf 0} & {\bf E}  \\ 
{\bf 0} &  {\bf M}_{y} & {\bf 0} \\
{\bf F} & {\bf 0}  &  {\bf L}_0  \end{pmatrix}. 
\end{eqnarray}
In writing Eq.~(\ref{eqT10}) we have neglected the synchrobetatron coupling effect that would be 
present if the RF cavity is located in a dispersive region~\cite{Huang.MatrixXZ.PRSTAB}. 
This should not impact the results below   
as it usually only causes a small correction.  
The usual dispersion decoupling matrix is
\begin{eqnarray}\label{eqUdef}
{\bf U} = 
\left( \begin{array}{ccc} {\bf I} & {\bf 0} & {\bf D}_{1}  \\ 
{\bf 0} &  {\bf I} & {\bf 0} \\
-{\bf D}_{1}^+ & {\bf 0}  &  {\bf I}  \end{array}\right),
\end{eqnarray}
where ${\bf D}_{1}=({\bf 0},{\bf d}_1)$, ${\bf d}_1=(D_1, D'_1)^T$, 
and the symplectic conjugate of matrix ${\bf D}_1$ is ${\bf D}_{1}^+= {\bf J}^T_2 {\bf D}_1^T {\bf J}_2$, with
\begin{eqnarray}
{\bf J}_2 = \left( \begin{array}{cc}  0  &  1 \\
   -1  & 0  \end{array} \right).
\end{eqnarray}
Applying the transformation ${\bf U}$ to ${\bf T}_1$, 
we get a new transfer matrix for the betatron coordinates 
${\bf X}=(x_\beta,x'_\beta,y,y',z,\delta)^T$
\begin{eqnarray}\label{eqT1nTotal}
{\bf T}_{1,n} &=& {\bf U}^{-1} {\bf T}_1 {\bf U}  \nonumber \\
 &=& {\bf T}_{1,n}^{(0)} + \epsilon \tilde{\bf T}_{1,n}, 
\end{eqnarray}
where $x_\beta=x-D_1\delta$, $x'_\beta=x'-D'_1\delta$, $ {\bf T}_{1,n}^{(0)}$ is a block-diagonal matrix,
\begin{eqnarray}
{\bf T}_{1,n}^{(0)} = 
\left( \begin{array}{ccc} {\bf M}_{x} & {\bf 0} & {\bf 0}  \\ 
{\bf 0} &  {\bf M}_y & {\bf 0} \\
{\bf 0} & {\bf 0}  &  {\bf L}  \end{array}\right)
\end{eqnarray}
 and 
\begin{eqnarray}\label{eqCplT1n}
\tilde{\bf T}_{1,n} &=& \left( \begin{array}{ccc} {\bf 0} & \tilde{\bf T}_{xy} & {\bf 0}  \\ 
\tilde{\bf T}_{yx} &  {\bf 0} & \tilde{\bf T}_{yz} \\
{\bf 0} &  \tilde{\bf T}_{zy} & {\bf 0} \\ \end{array}\right).
\end{eqnarray}
It has be shown that 
\begin{eqnarray}
\tilde{\bf T}_{xy} &=& -{\bf M}_{x,12}{\bf D}_2 {\bf W}{\bf M}_{y,21},  \label{eqTxy} \\
\tilde{\bf T}_{yx} &=& {\bf M}_{y,12}{\bf D}^T_2 ({\bf M}^{-1}_{x,21})^T {\bf J}_2, \label{eqTyx}  \\
\tilde{\bf T}_{yz} &=& {\bf M}_{y,12}{\bf W}{\bf L}_{21,n},  \label{eqTyz} \\
\tilde{\bf T}_{zy} &=& {\bf L}_{12,n}{\bf W}{\bf M}_{y,21}, \label{eqTzy}
\end{eqnarray}
where ${\bf L}_{21,n}$ is longitudinal transfer matrix from point 1 to 2 
with the (1,2) element replaced with $\bar{\eta}_{21}$ as defined in Eq. (19) of 
Ref.~\cite{Huang.MatrixXZ.PRSTAB} and likewise for ${\bf L}_{12,n}$.

From Eqs.~(\ref{eqCplT1n}-\ref{eqTzy}) it is seen that the longitudinal motion and the vertical motion 
are coupled through the vertical crab cavity via the off-diagonal blocks $\tilde{\bf T}_{yz}$ and $\tilde{\bf T}_{zy}$.
In addition, 
when the vertical crab cavity is located at 
a dispersive region (with nonzero horizontal dispersion ${\bf D}_2=(D_2, D'_2)^T$), 
the horizontal and vertical motion are also coupled through the crab cavity. 

The $x$-$y$ coupling and the $y$-$z$ coupling in Eq.~(\ref{eqT1nTotal}) can be 
simultaneously diagonalized. Analytic form of the decoupling transformation can be 
derived with a matrix perturbation approach. 
Let the transformation be denoted by the matrix ${\bf V}$, i.e., ${\bf V}^{-1}{\bf T}_{1,n}{\bf V}$ 
is block diagonal. Because the  matrix ${\bf T}_{1,n}$ deviates from the block diagonal matrix ${\bf T}_{1,n}^{(0)}$ by only a small 
amount that is proportional to $\epsilon$, we expect the deviation of $\bf V$ from the identity matrix 
to be proportional to $\epsilon$, too, i.e.,
\begin{eqnarray}\label{eqVgeneral}
{\bf V} = {\bf I} + \epsilon \tilde{\bf V}. 
\end{eqnarray}

We use a trial form of $\tilde{\bf V}$ 
\begin{eqnarray}\label{eqVtilde}
\tilde{\bf V} =
\left( \begin{array}{ccc} {\bf 0} & {\bf C}_{1}  & {\bf 0}   \\ 
-{\bf C}^+_1 & {\bf 0} &  {\bf C}_{2} \\
{\bf 0} &   -{\bf C}^+_2  &{\bf 0}  \end{array}\right).
\end{eqnarray}
It is easy to verify that the symplecticity of the matrix $\bf V$ is satisfied to first order of $\epsilon$ 
with $\tilde{\bf V}$ as given in Eq.~(\ref{eqVtilde}). 
The transfer matrix after applying the $\bf V$ transformation is 
\begin{eqnarray}
{\bf V}^{-1}{\bf T}_{1,n}{\bf V}  \approx  
({\bf I} - \epsilon \tilde{\bf V})({\bf T}_{1,n}^{(0)} + \epsilon \tilde{\bf T}_{1,n}) 
({\bf I} + \epsilon \tilde{\bf V}) \nonumber \\
= {\bf T}_{1,n}^{(0)} + \epsilon (\tilde{\bf T}_{1,n}-\tilde{\bf V}{\bf T}_{1,n}^{(0)}+{\bf T}_{1,n}^{(0)}\tilde{\bf V}) +
O(\epsilon^2). 
\end{eqnarray}
For the transfer matrix ${\bf V}^{-1}{\bf T}_{1,n}{\bf V}$ to be block diagonal to first order of $\epsilon$, 
we can require 
\begin{eqnarray}\label{eqtVdiag}
\tilde{\bf T}_{1,n}-\tilde{\bf V}{\bf T}_{1,n}^{(0)}+{\bf T}_{1,n}^{(0)}\tilde{\bf V} = 0, 
\end{eqnarray}
because the diagonal blocks of the l.h.s. of Eq.~(\ref{eqtVdiag}) are calculated to be all zeros. 
In fact, Eq.~(\ref{eqtVdiag}) is equivalent to 
\begin{eqnarray}
\tilde{T}_{xy}-{\bf C}_1 {\bf M}_y + {\bf M}_x {\bf C}_1 &=& 0, \label{eqTxyC1a} \\
\tilde{T}_{yx}+{\bf C}^+_1 {\bf M}_x - {\bf M}_y {\bf C}^+_1 &=& 0, \label{eqTxyC1b} \\
\tilde{T}_{yz}-{\bf C}_2 {\bf L} + {\bf M}_y {\bf C}_2 &=& 0, \label{eqTyzC2a}  \\
\tilde{T}_{zy}+{\bf C}^+_2 {\bf M}_y - {\bf L} {\bf C}^+_2 &=& 0. \label{eqTyzC2b}
\end{eqnarray}
From Eqs.~(\ref{eqTxyC1a}-\ref{eqTxyC1b}) one can solve for ${\bf C}_1$, and 
similarly from Eqs.~(\ref{eqTyzC2a}-\ref{eqTyzC2b}) for ${\bf C}_2$. The solutions are  
\begin{eqnarray}
\label{eqtC1form}
{\bf C}_1 &=& - \frac{\tilde{T}_{xy}+\tilde{T}_{yx}^+}{ {\rm Tr}({\bf M}_x-{\bf M}_y)},\\
\label{eqtC2form}
{\bf C}_2 &=& - \frac{\tilde{T}_{yz}+\tilde{T}_{zy}^+}{ {\rm Tr}({\bf M}_y-{\bf L})},
\end{eqnarray}
where $\rm Tr(\cdot)$ denotes taking the trace of a matrix. 
The solution for ${\bf C}_2$ would be the same if we had block diagonalized the $y$-$z$ plane only, 
ignoring the $x$-$y$ coupling in Eq.~(\ref{eqT1nTotal}). 
This indicates that the indirect $x$-$z$ coupling in Eq.~(\ref{eqT1nTotal}) is a second order 
effect. 
It is worth noting that Eqs.~(\ref{eqtC1form}-\ref{eqtC2form}) agree with the result of Ref.~\cite{SaganRubin.PRSTAB.2.074001} 
to first order of $\epsilon$. 

Inserting Eqs.~(\ref{eqTyz}-\ref{eqTzy}) into Eq.~(\ref{eqtC2form}),  and expressing the related vertical 
and longitudinal transfer matrices 
in terms of the beta functions and phase advances,
the four elements of 
\begin{eqnarray}
{\bf C}_2 &=& 
\left( \begin{array}{cc} C_{11} & C_{12} \\
C_{21} & C_{22}
  \end{array}\right)
\end{eqnarray}
 can be calculated and the results are
 \begin{widetext}
\begin{eqnarray}\label{eqC11full}
C_{11} &=&   \frac{\frac12 \epsilon \sqrt{\beta_1 \beta_2}}{{\cos2\pi\nu_s - \cos2\pi\nu_y}} 
\big(\cos\Psi_{s,12}\sin(2\pi\nu_y-\Psi_{12})+
 \cos(2\pi\nu_s-\Psi_{s,12})\sin\Psi_{12} \big), \\ \label{eqC12full}
C_{12} &=&   \frac{\frac12 \epsilon\beta_s\sqrt{\beta_1 \beta_2}}{{\cos2\pi\nu_s - \cos2\pi\nu_y}} 
\big(\sin\Psi_{s,12}\sin(2\pi\nu_y-\Psi_{12})- 
 \sin(2\pi\nu_s-\Psi_{s,12})\sin\Psi_{12} \big), \\ 
C_{21} &=&   \frac{\frac12\epsilon\sqrt{\beta_2/ \beta_1}}{{\cos2\pi\nu_s - \cos2\pi\nu_y}} 
\big(
\cos(2\pi\nu_s-\Psi_{s,12})(\cos\Psi_{12}-\alpha_1 \sin\Psi_{12}) 
-\cos\Psi_{s,12}(\cos(2\pi\nu_y-\Psi_{12})+\alpha_1 \sin(2\pi\nu_y-\Psi_{12})) 
 \big), \nonumber \\  \label{eqC21full}
 \\
C_{22} &=&   \frac{\frac12\epsilon\beta_s\sqrt{\beta_2/ \beta_1}}{{\cos2\pi\nu_s - \cos2\pi\nu_y}} 
\big(
\sin(2\pi\nu_s-\Psi_{s,12})(\cos\Psi_{12}-\alpha_1 \sin\Psi_{12}) 
+\sin\Psi_{s,12}(\cos(2\pi\nu_y-\Psi_{12})+\alpha_1 \sin(2\pi\nu_y-\Psi_{12})) 
 \big), \nonumber \\   \label{eqC22full}
\end{eqnarray}  
\end{widetext}
where 
$\Psi_{s,12}$ is the synchrotron phase advance from point 2 to 1, $\beta_s$ is the longitudinal 
beta function, 
$\nu_s$ is the synchrotron tune, $\nu_y$ is the vertical tune,
$\alpha_1$ and $\beta_{1}$ are the vertical Courant-Snyder functions at point 1, 
$\Psi_{12}$ is the vertical betatron phase 
advance from point 2 to 1, 
and $\beta_2$ is the vertical beta function at point 2.
Because the longitudinal motion is slow,  
it can be ignored to simplify the results.  
The results under this assumption can be obtained from the exact 
formulas by using the approximations $\cos 2\pi\nu_s \approx 1$ and $\beta_s \sin 2\pi\nu_s \approx \bar{\eta}$,
with  $\bar{\eta}=-\oint D/\rho ds$. 
The simplified expressions for the ${\bf C}_2$ matrix elements are found to be 
\begin{eqnarray}\label{eqC11}
C_{11} &=&  \epsilon \frac{\sqrt{\beta_1 \beta_2}}{{2\sin \pi\nu_y}} {\cos(\pi\nu_y - \Psi_{12})},  \\  \label{eqC21}
C_{12} &=& \epsilon \frac{\bar{\eta}\sqrt{\beta_1 {\beta_2}}}{2\sin \pi \nu_y}
\big[\frac{\sin \Psi_{12}}{2\sin \pi \nu_y} - \frac{\bar{\eta}_{12}}{\bar{\eta}} \cos (\pi\nu_y - \Psi_{12} ) \big],\\
C_{21} &=& \epsilon \frac{\sqrt{\beta_2/{\beta_1}}}{2\sin \pi\nu_y}\left[{\sin(\pi\nu_y - \Psi_{12})}
-\alpha_1 \cos (\pi\nu_y - \Psi_{12})\right], \nonumber \\ \\
C_{22} &=& \epsilon \frac{\bar{\eta}\sqrt{\beta_2/{\beta_1}}}{2\sin \pi\nu_y}
\big[ \frac1{2\sin \pi \nu_y}(\cos \Psi_{12}-\alpha_1 \sin \Psi_{12})  \nonumber \\
& &  -\frac{\bar{\eta}_{12}}{\bar{\eta}} (\sin (\pi\nu_y - \Psi_{12}) - \alpha_1 \cos (\pi\nu_y - \Psi_{12} ))
\big],  \label{eqC22}
\end{eqnarray}    
It is  worth noting that 
\begin{eqnarray}
||{\bf C}_2|| &=& \frac18  \frac{\epsilon^2 \beta_2 \bar{\eta}}{\sin^2 \pi \nu_y \tan \pi \nu_y}, 
\end{eqnarray}
which is a constant all around the ring. 

In the above we showed that a vertical crab cavity causes $y$-$z$ coupling, and additionally 
$x$-$y$ coupling if it is located 
at a dispersive region. The coupled motioned can be decoupled with a linear transformation.  
The transformation for the $y$-$z$ coupling is given by Eqs.~(\ref{eqVgeneral}),(\ref{eqVtilde}), 
and (\ref{eqC11}-\ref{eqC22}). 
 
\section{\label{secYZOhmi}Equilibrium distribution in an electron storage ring with crab cavity}
\subsection{Beam distribution changes due to a crab cavity}
The decoupled coordinates ${\bf X}_d$ are related to the original coordinates 
${\bf X} $ through ${\bf X} = {\bf UV}{\bf X}_d$. 
The second order moment matrices of a particle distribution in ${\bf X}$ and ${\bf X}_d$ 
coordinates, defined as
\begin{eqnarray}
{\bf \Sigma} = <{\bf X}{\bf X}^T>, \quad {\bf \Sigma}_d = <{\bf X}_d{\bf X}_d^T>,
\end{eqnarray}
 are related through
\begin{eqnarray}\label{eqSigUV}
{\bf \Sigma} = {\bf UV}{\bf \Sigma_d}{\bf V}^T{\bf U}^T,
\end{eqnarray}
where $<\cdot>$ denotes taking average over the particle distribution and we have assumed the 
distribution is centered on the reference orbit. 
The moment matrices are symmetric. 
Matrix ${\bf \Sigma_d}$ is block diagonal. The two matrices ${\bf \Sigma}$ and ${\bf \Sigma}_d$ 
may be written as
\begin{eqnarray}\label{eqSigma}
{\bf \Sigma} = \begin{pmatrix}
{\bf \Sigma}_{xx} & {\bf \Sigma}_{xy} & {\bf \Sigma}_{xz} \\
{\bf \Sigma}^T_{xy} & {\bf \Sigma}_{yy} & {\bf \Sigma}_{yz} \\
{\bf \Sigma}^T_{xz}  & {\bf \Sigma}^T_{yz}  & {\bf \Sigma}_{zz}
\end{pmatrix}
\end{eqnarray}
and 
\begin{eqnarray}\label{eqSigmad}
{\bf \Sigma}_d = \begin{pmatrix}
{\bf \Sigma}_{x} & {\bf 0} & {\bf 0} \\
{\bf 0} & {\bf \Sigma}_{y} & {\bf 0} \\
{\bf 0}  & {\bf 0}  & {\bf \Sigma}_{z}
\end{pmatrix}.
\end{eqnarray}
Inserting Eqs.~(\ref{eqUdef},\ref{eqVgeneral}-\ref{eqVtilde}) into Eq.~\ref{eqSigUV}, 
the block matrices in Eqs.~(\ref{eqSigma}-\ref{eqSigmad}) are related. 
It is found that the changes to the diagonal blocks of ${\bf \Sigma}$ due to the crab cavity 
are second order effects, i.e., of the order $O(\epsilon^2)$, for example
\begin{eqnarray}\label{eqSigmayy1}
{\bf \Sigma}_{yy}&=& {\bf C}^+_1{\bf \Sigma}_{x}({\bf C}^+_1)^T+{\bf \Sigma}_{y}+
{\bf C}_2{\bf \Sigma}_{z}{\bf C}^T_2.
\end{eqnarray}
Therefore, it may be inferred that the deviation of the diagonal block matrices in ${\bf \Sigma}_d$ 
from the case when the crab cavity is off (e.g., assuming it was adiabatically turned on) 
is also a second order effect. 
In other words, we assume
\begin{eqnarray}
{\bf \Sigma}_{x} &\approx& {\bf \Sigma}_{x0}=\epsilon_x \begin{pmatrix} \beta_x & -\alpha_x \\ -\alpha_x & \gamma_x \end{pmatrix}, \\
{\bf \Sigma}_{y} &\approx& {\bf \Sigma}_{y0}=\epsilon_y \begin{pmatrix} \beta_y & -\alpha_y \\ -\alpha_y & \gamma_y \end{pmatrix}, \\
{\bf \Sigma}_{z} &\approx& {\bf \Sigma}_{z0}= \begin{pmatrix} \sigma_z^2 & 0 \\ 0 & \sigma_\delta^2 \end{pmatrix}, \label{eqSigZ0tmp}
\end{eqnarray}
where $\epsilon_{x,y}$ are original horizontal and vertical emittances, 
$\alpha_{x,y}$, $\beta_{x,y}$, and $\gamma_{x,y}$ are the
Courant-Snyder functions for the horizontal and vertical directions, with $\gamma_{x,y}=(1+\alpha_{x,y}^2)/\beta_{x,y}$, and
$\sigma_z$ and $\sigma_\delta$ are original bunch length and momentum spread, respectively.  
This assumption is  validated in the next subsection with the Ohmi envelope approach. 
In Eq. (\ref{eqSigZ0tmp}) we have assumed $\alpha_s=0$, i.e., there is no tilt between the $z$-$\delta$ directions. 
The deviation of matrix block ${\bf \Sigma}_{xz}$ from the original case without crab cavity is also of the second order.
Only  ${\bf \Sigma}_{xy}$ and  ${\bf \Sigma}_{yz}$ have first order dependence over the crab cavity strength parameter $\epsilon$, 
\begin{eqnarray}
\label{eqSigmaxy} 
{\bf \Sigma}_{xy}&=& -\Sigma_x ({\bf C}^+_1)^T+({\bf C}_1-{\bf D}_1{\bf C}_2^+){\bf \Sigma}_y+{\bf D}_1{\bf \Sigma}_z{\bf C}_2^T, \\
\label{eqSigmayz}
{\bf \Sigma}_{yz} &=& {\bf C}^+_1{\bf \Sigma}_{x}({\bf D}^+_1)^T-{\bf \Sigma}_{y}[({\bf C}^+_2)^T+{\bf C}^T_1({\bf D}^+_1)^T]  \nonumber \\
&& +{\bf C}_2{\bf \Sigma}_{z}.
\end{eqnarray} 

We are interested in the tilt across the $y$-$z$ planes introduced by the crab cavity. 
In an electron storage ring,  if originally there is no horizontal to vertical coupling, typically 
${\bf \Sigma}_{y0}\approx0$. Also, because the horizontal emittance is typically much smaller than 
the longitudinal emittance, unless the horizontal and vertical motions are near a resonance, normally the contribution from 
the ${\bf \Sigma}_{x}$ term in Eq.~(\ref{eqSigmayz}) is much smaller than the last term. 
Keeping only the last term, we obtain  
\begin{eqnarray}\label{eqSigmayz2}
{\bf \Sigma}_{yz}&=& \begin{pmatrix} \sigma_{yz} & \sigma_{y\delta} \\ \sigma_{y'z} & \sigma_{y'\delta} \end{pmatrix}
 \approx {\bf C}_2{\bf \Sigma}_{z}=\begin{pmatrix} C_{11} \sigma_z^2 & C_{12}\sigma_\delta^2
 \\ C_{21}\sigma_z^2 & C_{22}\sigma_\delta^2 \end{pmatrix}.
\end{eqnarray}

From Eq. (\ref{eqSigmayz2}) we see that the crab cavity causes a tilt of the beam distribution 
between the vertical and longitudinal directions. The tilt is not only between the vertical coordinates 
and the $z$-coordinate, but also the $\delta$-coordinate.

\subsection{First order perturbation to the equilibrium distribution by a crab cavity}
In an electron storage ring, the beam reaches an equilibrium distribution determined by 
the balance between quantum excitation and radiation damping. 
The equilibrium distribution at a location of the ring can be found by 
solving Ohmi's envelope equation~\cite{Ohmi.PRE.49.751},
\begin{eqnarray}\label{eqOhmi0}
{\bf T}_0{\bf \Sigma}_0{\bf T}^T_0 + \bar{\bf B}_0 = {\bf \Sigma}_0,
\end{eqnarray}
where ${\bf T}_0$ is the one-turn transfer matrix (including damping), ${\bf \Sigma}_0$ is 
the second order moment matrix as defined in Eq.~(\ref{eqSigma}), and $\bar{\bf B}_0$ is 
the one-turn integrated diffusion matrix
\begin{eqnarray}\label{eqOhmiB0}
{\bf B}_0(s_0) &=& \int_{s_0}^{s_0+C} T_{s_0+C, s'} B(s') T_{s_0+C, s'}^T ds',
\end{eqnarray}
where $T_{s_0+C, s'}$ is the transfer matrix from $s'$ to $s_0+C$, $C$ is the ring circumference, 
and $B(s')$ is the diffusion matrix at location $s=s'$.
In Eqs.~(\ref{eqOhmi0}-\ref{eqOhmiB0})
subscript $0$ indicates the case without the crab cavity. 

When the crab cavity is introduced to the ring, all quantities in Eq.~(\ref{eqOhmi0}) are 
changed. Suppose we are concerned of a point immediately downstream of the 
crab cavity, the envelope equation becomes
\begin{eqnarray}\label{eqOhmiC}
{\bf T}{\bf \Sigma}{\bf T}^T + \bar{\bf B} = {\bf \Sigma},
\end{eqnarray}
with the new one-turn transfer matrix and new integrated diffusion matrix being
\begin{eqnarray}
{\bf T}&=&{\bf T}_c{\bf T}_0, \\
\bar{\bf B}&=& {\bf T}_c\bar{\bf B}_0{\bf T}_c^T. 
\end{eqnarray}
Multiplying ${\bf T}_c^{-1}$ and $({\bf T}_c^T)^{-1}$ from the left and right sides to Eq.~(\ref{eqOhmiC}), respectively, 
and inserting $\bar{\bf B}_0$ from Eq.~(\ref{eqOhmi0}), we get
\begin{eqnarray}\label{eqOhmitmp}
{\bf T}_0({\bf \Sigma}-{\bf \Sigma}_0){\bf T}_0^T  = {\bf T}_c^{-1}{\bf \Sigma}({\bf T}_c^T)^{-1}-{\bf \Sigma}_0. 
\end{eqnarray}
Rewriting ${\bf T}_c$ from Eq.~(\ref{eqTcgen}) as 
\begin{eqnarray}
{\bf T}_c = {\bf I} + \epsilon \tilde{\bf W}, \quad 
\tilde{\bf W} = \begin{pmatrix} {\bf 0} & {\bf 0} & {\bf 0} \\ {\bf 0} & {\bf 0} & {\bf W} \\
{\bf 0} & {\bf W} & {\bf 0}
\end{pmatrix}.
\end{eqnarray}
The inverse matrices of ${\bf T}_c$ and its transpose are
\begin{eqnarray}
{\bf T}_c^{-1} &=& {\bf I} - \epsilon \tilde{\bf W}, \\
({\bf T}_c^T)^{-1} &=& {\bf I} - \epsilon \tilde{\bf W}^T,
\end{eqnarray}
with which Eq.~(\ref{eqOhmitmp}) becomes
\begin{eqnarray}\label{eqOhmiD}
{\bf T}_0 {\bf \Delta}{\bf T}_0^T  - {\bf \Delta}= -\epsilon (\tilde{\bf W}{\bf \Sigma}+{\bf \Sigma}\tilde{\bf W}^T)
+\epsilon^2 (\tilde{\bf W}{\bf \Sigma}\tilde{\bf W}^T),  \nonumber \\
\end{eqnarray}
where we have used the definition
\begin{eqnarray}
{\bf \Delta}&=& {\bf \Sigma}-{\bf \Sigma}_0.
 = \begin{pmatrix}
{\bf \Delta}_{xx} & {\bf \Delta}_{xy} & {\bf \Delta}_{xz} \\
{\bf \Delta}^T_{xy} & {\bf \Delta}_{yy} & {\bf \Delta}_{yz} \\
{\bf \Delta}^T_{xz}  & {\bf \Delta}^T_{yz}  & {\bf \Delta}_{zz}
\end{pmatrix}. 
\end{eqnarray}
Changes of the 
equilibrium distribution caused by the crab cavity can be found by solving Eq.~(\ref{eqOhmiD}) for ${\bf \Delta}$.
 
For results to first order of 
the strength parameter $\epsilon$,  on the r.h.s. of Eq.~(\ref{eqOhmiD}) 
${\bf \Sigma}$ can be replaced by
the original second order moment matrix, ${\bf \Sigma}_0={\bf U}{\bf \Sigma}_{d0}{\bf U}^T$, 
with ${\bf \Sigma}_{d0}$ a block diagonal matrix as in Eq.~(\ref{eqSigmad}), and 
the $\epsilon^2$ term can be dropped. 
In this case, among the $2\times2$ sub-blocks of the r.h.s of Eq.~(\ref{eqOhmiD}), only 
the $x-y$ and $y-z$ blocks and their symmetric counterparts are nonzero. 
Therefore the elements in the sub blocks ${\bf \Delta}_{xx}$, ${\bf \Delta}_{yy}$, ${\bf \Delta}_{zz}$, and 
${\bf \Delta}_{xz}$ are solutions of a linear homogeneous equation set.
In general this equation set is non-degenerate (the ${\bf T}_0$ matrix includes damping). 
Therefore these blocks are all zeros to the first order of 
$\epsilon$, which verifies the assumption we made in the previous subsection.

When the crab cavity is located in a dispersion region, 
the elements of ${\bf \Delta}_{xy}$ and ${\bf \Delta}_{yz}$ blocks are coupled in an inhomogeneous linear equation 
set, 
\begin{eqnarray}\label{eqDeltaxy}
{\bf M}_y{\bf \Delta}_{xy}^T{\bf M}_{x}^T+{\bf M}_y{\bf \Delta}_{yz}{\bf E}^T-{\bf \Delta}_{xy}^T = \phantom{i + j+j + k + k} \nonumber \\
  (-\epsilon) \left( -{\bf \Sigma}_{x0}({\bf D}_1^+)^T{\bf W}^T +{\bf D}_1 {\bf \Sigma}_{z0}{\bf W}^T \right), \phantom{i+j+j+ } \\
{\bf M}_x{\bf \Delta}_{xy}^T{\bf F}^T+{\bf M}_y{\bf \Delta}_{yz}{\bf L}^T-{\bf \Delta}_{yz} = \phantom{i + j a+ k+ j+j } \nonumber \\
 (-\epsilon) \left( {\bf W}{\bf \Sigma}_{z0}+{\bf W}{\bf D}_1^+{\bf \Sigma}_{x0}({\bf D}_1^+)^T+{\bf \Sigma}_{y0}{\bf W}^T \right),
 \phantom{i +j } 
  \label{eqDeltayz}
\end{eqnarray}
where ${\bf \Sigma}_{x0}$, ${\bf \Sigma}_{y0}$, and ${\bf \Sigma}_{z0}$ are diagonal blocks of ${\bf \Sigma}_{d0}$.
For electron storage rings initially without $x$-$y$ coupling, the original vertical emittance is zero and hence ${\bf \Sigma}_{y0}=0$. 
Eqs.(\ref{eqDeltaxy}-\ref{eqDeltayz}) can be solved for elements of ${\bf \Delta}_{xy}$ and ${\bf \Delta}_{yz}$. 
The coupling terms due to dispersion are on the order of $O(\mathcal{H}_x/\beta_s)$, which is usually very small, 
where $\mathcal{H_x}=(D^2+(\alpha_x D+\beta_x D')^2)/2\beta_x$ is the dispersion invariant. 
For example, SPEAR3 has $\beta_s \approx 6.2$~m and at the standard straight sections $\mathcal{H}=1.0$~mm. 
Ignoring the coupling terms, the solution for elements of ${\bf \Delta}_{yz}$ is, 
\begin{eqnarray}
{\bf \Sigma}_{yz}&=&{\bf \Delta}_{yz} ={\bf C}_2{\bf \Sigma}_{z0} \nonumber \\
&\approx&  \frac{\epsilon\sigma_z^2}{2\sin\pi\nu_y} \begin{pmatrix}
\beta_y \cos \pi \nu_y & 0 \\
\sin \pi \nu_y-\alpha_y \cos \pi \nu_y & \frac{\bar{\eta} \gamma_s^2}{2\sin\pi\nu_y}
\end{pmatrix}, 
\end{eqnarray}
where $\gamma_s =\sigma_z/\sigma_\delta$. 
This result is the same as given by 
Eqs. (\ref{eqSigmayz2},\ref{eqC11}-\ref{eqC22}) for the location just downstream of the crab cavity.

\subsection{Vertical eigen-emittance due to crab cavity}
When solving the Ohmi envelope equation, Eq. (\ref{eqOhmiC}), for the equilibrium beam distribution with first order perturbation, 
we found that normal mode distributions don't change. 
However, because the longitudinal dimension of bunched beams in storage rings is usually much larger than the transverse dimensions, 
second order terms involving the longitudinal dimension may also be important. 
In fact, numeric solutions of  Eq. (\ref{eqOhmiC}) show that the normal mode distributions 
do have changes of order $\epsilon^2$. 
Notably, there is a finite vertical normal mode emittance (i.e., eigen-emittance) and the bunch length 
changes. 

Because coupling with the horizontal direction is small, we can consider only the vertical and longitudinal 
directions. Assuming the original vertical emittance is zero, ${\bf \Sigma}_{y0}=0$, 
and using  
\begin{eqnarray}\label{eqDeltaElems}
{\bf \Delta}_{zz} = \begin{pmatrix}\Delta \sigma_{zz} & \Delta \sigma_{z\delta} \\ \Delta \sigma_{z\delta} & \Delta \sigma_{\delta\delta} 
\end{pmatrix}, \quad
{\bf \Delta}_{yz} = \begin{pmatrix}\sigma_{yz} &  \sigma_{y\delta} \\  \sigma_{y'z} &  \sigma_{y'\delta}
\end{pmatrix}, \nonumber \\
{\bf \Delta}_{yy} = \begin{pmatrix}\sigma_{yy} &  \sigma_{yy'} \\  \sigma_{yy'} &  \sigma_{y'y'}
\end{pmatrix},  \phantom{adflakjdflakj}
\end{eqnarray}
the coupled matrix equations from Eq. (\ref{eqOhmiD}) are given by
 \begin{eqnarray}
{\bf L} {\bf \Delta}_{zz} {\bf L}^T-{\bf \Delta}_{zz}&=&
    (-\epsilon)\begin{pmatrix}0 &  \sigma_{yz} \\  \sigma_{yz} &  2\sigma_{y\delta}-\epsilon \sigma_{yy}
\end{pmatrix}
 \label{eqDeltazz2a} \\
{\bf M}_{y} {\bf \Delta}_{yz} {\bf L}^T-{\bf \Delta}_{yz}&=& 
   (-\epsilon)\begin{pmatrix}0 &  \sigma_{yy} \\  \sigma_{zz} &  \sigma_{z\delta}+\sigma_{yy'}-\epsilon \sigma_{yz}
\end{pmatrix}  \label{eqDeltayz2a} \\
{\bf M}_{y} {\bf \Delta}_{yy} {\bf M}_y^T-{\bf \Delta}_{yy} &=& 
     (-\epsilon)\begin{pmatrix}0 &  \sigma_{yz} \\  \sigma_{yz} &  2\sigma_{y'z}-\epsilon \sigma_{zz}
\end{pmatrix},  \label{eqDeltayy2a} 
\end{eqnarray}
where $\sigma_{zz}=\sigma_z^2=\sigma_{z0}^2+\Delta \sigma_{zz}$, 
$\sigma_{z\delta}=\Delta \sigma_{z\delta}$, and $\sigma_{z0}$ is the original bunch length. 

In principle, solving Eqs. (\ref{eqDeltazz2a}-\ref{eqDeltayy2a}) for 
matrix elements in Eq. (\ref{eqDeltaElems}) gives the equilibrium distribution, from which one 
can calculate the eigen-emittances. 
Although the general solution has not been found, approximate results can be derived from these equations. 
From Eqs. ( \ref{eqDeltazz2a}) and (\ref{eqDeltayz2a}) we obtain
\begin{eqnarray}
\Delta \sigma_{zz}- \beta_s^2 \Delta \sigma_{\delta\delta} = \sigma_{yz} \epsilon \beta_s \cot \psi_s, \\
\sigma_{yz} \approx \sigma_{zz} \frac{\epsilon}2 \frac{\beta_y}{\cos\psi_s-\cos\psi_y}.
\end{eqnarray}
Numeric solutions indicate that $\beta_s^2 \Delta \sigma_{\delta\delta}\ll \Delta \sigma_{zz}$. Ignoring 
the $\beta_s^2 \Delta \sigma_{\delta\delta}$ term, we obtain
\begin{eqnarray}\label{eqApproxSigz}
\sigma_z&\approx& \sigma_{z0} \big(1-\frac{\epsilon^2}2\beta_s\beta_y \cot\psi_s \frac{\sin\psi_y}{\cos\psi_s-\cos\psi_y} \big)^{-1/2}.
\end{eqnarray}
The projected distribution on the ($z$, $\delta$) plane, ${\bf \Sigma}_{zz}$, and the longitudinal 
normal mode distribution, ${\bf \Sigma}_{z}$, are approximately equal (their difference is 
a small term $\propto \epsilon^2 \epsilon_y$, where $\epsilon_y$ is the vertical eigen-emittance). 
Therefore, the longitudinal eigen-emittance is 
\begin{eqnarray}
\epsilon_{z}&\approx& \sigma_z \sigma_{\delta 0},
\end{eqnarray}
where $\sigma_z$ is given in Eq. (\ref{eqApproxSigz}) and $\sigma_{\delta 0}$ is the original 
momentum spread. 

A different approach has been taken to obtain the vertical eigen-emittance due to the crab cavity. 
In an electron storage ring, the source of finite emittances is the stochastic photon emission in 
magnetic fields. 
The crab cavity causes a tilt across the longitudinal and vertical planes. Consequently, 
the energy loss of an electron due to photon emission will cause a random shift of vertical coordinates 
relative to its ''ideal orbit'', which gives rise to the vertical eigen-emittance.
 
From Eq. (\ref{eqSigmayz2}) we find the correlation of the $y$, $y'$ coordinates with momentum 
deviation $\delta$,
\begin{eqnarray}
r_{y\delta} &=& \frac{\sigma_{y\delta}}{\sigma_y \sigma_\delta}=C_{12}, 
\qquad
r_{y'\delta} = \frac{\sigma_{y'\delta}}{\sigma_{y'} \sigma_\delta}=C_{22}.
\end{eqnarray} 
The equivalent vertical coordinate displacements for an energy loss $\Delta \delta$ are thus
\begin{eqnarray}
\Delta y &=& C_{12}\Delta \delta, \qquad
\Delta y'= C_{22}\Delta \delta,
\end{eqnarray} 
and the quandratic term of the Courant-Snyder invariant change is
\begin{eqnarray}
\Delta J_y &=&  \mathcal{H}_c\Delta \delta^2,
\end{eqnarray} 
where we defined  crab cavity dispersion invariant 
\begin{eqnarray}\label{eqHcdef}
\mathcal{H}_c &=&  \frac1{\beta_y}\big( C_{12}^2+(\alpha_y C_{12}+\beta_y C_{22})^2\big). 
\end{eqnarray} 
The increase of vertical emittance due to photon emission in a tilted bunch is the same as 
due to vertical dispersion, except here the vertical dispersion invariant is replaced with 
the quantity $\mathcal{H}_c$. 

Using Eqs. (\ref{eqC12full}, \ref{eqC22full}), and integrating $\mathcal{H}_c$ over the ring, we 
obtain the average  crab cavity dispersion invariant
\begin{eqnarray}\label{eqHcAvg}
<\mathcal{H}_c> &=&  \frac{\epsilon^2\bar{\eta}^2\beta_2}{12}
\frac{2+\cos2\pi\nu_y}{(\cos2\pi\nu_s-\cos2\pi\nu_y)^2},
\end{eqnarray} 
where we have assumed the synchrotron phase advance is linearly proportional to distance traveled in 
bending magnets. 
Similar to vertical emittance due to the usual vertical dispersion~\cite{Lee_AP}, the vertical eigen-emittance for an 
isomagnetic storage ring due to crab cavity is given by
\begin{eqnarray}\label{eqEigenEmitY}
\epsilon_{y} &=& C_q \frac{\gamma^2 <\mathcal{H}_c> }{J_y \rho},
\end{eqnarray} 
where $C_q=3.83\times10^{-13}$~m, $\gamma$ is the Lorentz energy factor, $J_y=1$ is the vertical damping 
partition, and $\rho$ is bending radius. 

\subsection{Prediction of short pulse performance}
For the crab cavity application of generating short pulses, 
an important task is to estimate the expected short pulse performance, such 
as the minimum pulse duration and the fraction of flux accepted by a slit with certain 
aperture. This can be done if we know the beam distribution at the source point and the photon beam 
optics between the source point and the slit. 
 
\begin{figure}[hbt]
\psfrag{z}{\small{$z$}}
\psfrag{yp}{\small{$y'$}}
\psfrag{sig1}{\small{$\sigma_z$}}
\psfrag{sig2}{\small{$\sigma_{y'}$}}
\psfrag{width}{\small{$\sigma_{zm}=\sigma_z \sqrt{1-\frac{\sigma_{y'z}^2}{\sigma_z^2\sigma_{y'}^2} }$}}
  \centering
  \includegraphics[width=2.8in]{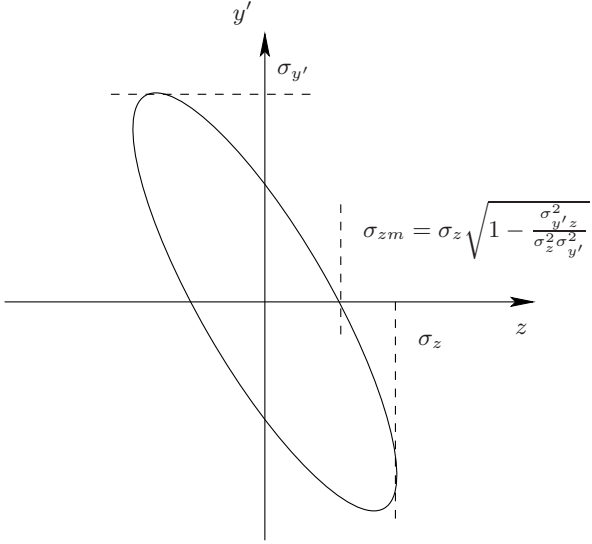}
  \caption{\label{figEllipse} The projected $y'$-$z$ ellipse.  
  }
\end{figure}

The normal mode distributions for the longitudinal and vertical planes are Gaussian. 
Their projections onto the $y$-$z$ plane or 
$y'$-$z$ plane are hence also Gaussian. 
Because the slit is usually placed far away from the source point, the tilt of the photon beam 
is primarily determined by the $y'$-$z$ tilt at the source point. 
The distribution function for the $y'$-$z$ projection can be written as  
\begin{eqnarray}
\rho(y',z) &=& \frac1{2\pi \epsilon_{y'z}} \exp\big(-\frac{\sigma_{z}^2y^2-2\sigma_{y'z}y'z+\sigma_{y'}^2z^2}{2 \epsilon_{y'z}^2 } \big),
\end{eqnarray}
where  $ \epsilon_{y'z}=\sqrt{\sigma_z^2\sigma_{y'}^2-\sigma_{y'z}}$ is the projected emittance. An ellipse that represents the $y'$-$z$ distribution 
is shown in FIG.~\ref{figEllipse}. The intercept with the $z$-axis signifies the minimum bunch length. It is given by 
\begin{eqnarray}
\sigma_{zm} &=&\frac{\epsilon_{y'z}}{\sigma_{y'}}=\sigma_z \sqrt{1-\frac{\sigma_{y'z}^2}{\sigma_z^2\sigma_{y'}^2} }.
\end{eqnarray}

With a finite vertical eigen-emittance, the $y$-$z$ and $y$-$y$ blocks of the sigma matrix are 
related to the normal mode distribution through
\begin{eqnarray}\label{eqSigmayz3}
{\bf \Sigma}_{yz}&=& {\bf C}_2 {\bf \Sigma}_{z} - {\bf \Sigma}_y ({\bf C}_2^+)^T, \\
{\bf \Sigma}_{yy}&=& {\bf C}_2 {\bf \Sigma}_{z}{\bf C}_2^T + {\bf \Sigma}_y. \label{eqSigmayy3}
\end{eqnarray} 
For simplicity we assume that at the source point $\alpha_y=0$, as this is usually the case. 
From Eqs. (\ref{eqSigmayz3}-\ref{eqSigmayy3}), we obtain  
\begin{eqnarray}\label{eqsigmaypz2}
\sigma_{y'z} &=& C_{21}\sigma_z^2+C_{12}\frac{\epsilon_y}{\beta_y}, \\
\sigma_{y'y'} &=& C_{21}^2 \sigma_z^2 + C_{22}^2\sigma_\delta^2+\frac{\epsilon_y}{\beta_y}+\sigma_\theta^2, \label{eqsigmaypyp2}
\end{eqnarray}
where we have added the contribution of radiation divergence, $\sigma_\theta$, to $\sigma_{y'y'}$, 
such that Eqs. (\ref{eqsigmaypz2}-\ref{eqsigmaypyp2}) are for the photon beam distribution at the source point. 

Ignoring small terms, the projected emittance on the $y'$-$z$ plane is given by
\begin{eqnarray}\label{eqepsilonypz}
\epsilon_{y'z} &=& \sigma_{z} \big( C_{22}^2\sigma_\delta^2+\frac{\epsilon_y}{\beta_y}+\sigma_\theta^2 \big)^{1/2}.
\end{eqnarray}
The minimum pulse duration is thus
\begin{eqnarray}\label{eqMinSigt}
\sigma_{zm} &=& \sigma_{z} \Big( \frac{ C_{22}^2\sigma_\delta^2+\frac{\epsilon_y}{\beta_y}+\sigma_\theta^2}
{ C_{21}^2 \sigma_z^2 +C_{22}^2\sigma_\delta^2+\frac{\epsilon_y}{\beta_y}+\sigma_\theta^2} \Big)^{1/2} \nonumber \\
&\approx& \frac1{C_{21}}\sqrt{\frac{\epsilon_y}{\beta_y}+\sigma_\theta^2},
\end{eqnarray}
where the approximate equality is valid  when the $\sigma_\delta^2$ term is negligible. 
For hard X-ray sources, the radiation divergence term may also be negligible. 
In this case, if there is no other sources of vertical emittance,
the minimum pulse duration is independent of the strength of the crab cavity. 
With a low deflecting voltage, the $y$-$z$ tilt of the photon beam at the slit is small, which requires 
a small slit aperture for a given accepted flux. 
Of course, in reality a reasonable deflecting voltage is needed to overcome the contribution of 
the finite vertical emittance due to spurious vertical 
dispersion and horizontal to vertical coupling and the finite radiation divergence. 
An optimal deflecting voltage is probably achieved when the term of the crab cavity induced eigen-emittance is 
a few times of the contributions of original vertical emittance and radiation divergence. 
The longitudinal distribution of the short pulse accepted by a given slit 
aperture is calculated from the distribution function 
$\rho(y',z) $ with 
\begin{eqnarray}\label{eqFlux}
\lambda(z;y_a) &=& \int_{-y_a/L_a}^{y_a/L_a} dy' \rho(y',z),
\end{eqnarray}
where $y_a$ is the half aperture, $L_a$ is the distance from the slit to the source point, and
we have ignored the finite vertical size of the electron beam at the source point. 
The percentage of total flux in the accepted pulse can be calculated with 
\begin{eqnarray}\label{eqFluxpct}
F(y_a) &=& \int dz \lambda(z;y_a),
\end{eqnarray}
and the pulse duration $\sigma_{zp}$ can be obtained from
\begin{eqnarray}\label{eqPulseLen}
\sigma_{zp}^2(y_a) &=& \frac{\int dz z^2 \lambda(z;y_a)}{\int dz \lambda(z;y_a)}.
\end{eqnarray}

\section{\label{secXcrab}Coupled motion by a horizontal crab cavity}
Because of the lack of vertical dispersion, 
the coupling due to a horizontal crab cavity generally does not 
involve the vertical plane. 
Therefore we only need to study the 4D phase space coordinates, $(x,x',z,\delta)$. 
The transfer matrix for the crab cavity is 
\begin{eqnarray}\label{eqXTcgen}
{\bf T}_{c} = {\bf I}+ \epsilon {\bf W}_x, \quad \tilde{\bf W}_x = 
\left( \begin{array}{ccc}  {\bf 0} & {\bf W}  \\ 
{\bf W} &   {\bf 0}  \end{array}\right).
\end{eqnarray}
Assuming the crab cavity is located at point 2, the one-turn transfer matrix at point 1 is
\begin{eqnarray}
{\bf T}_1 &= & {\bf T}_{12}{\bf T}_{c}{\bf T}_{21} = {\bf T}_1^{(0)} + \epsilon \tilde{\bf T}_1, 
\end{eqnarray}
with 
\begin{eqnarray}
\tilde{\bf T}_1 = \phantom{some space to shift it to the left. More space  space} \nonumber \\
 \begin{pmatrix} {\bf E}_{12}{\bf W}{\bf M}_{21}+{\bf M}_{12}{\bf W}{\bf F}_{21} 
& {\bf E}_{12}{\bf W}{\bf E}_{21} +{\bf M}_{12}{\bf W}{\bf L}_{21}  \\
 {\bf L}_{12}{\bf W}{\bf M}_{21}+{\bf F}_{12}{\bf W}{\bf F}_{21} 
& {\bf L}_{12}{\bf W}{\bf E}_{21} +{\bf F}_{12}{\bf W}{\bf F}_{21}  
  \end{pmatrix}, \nonumber \\
\end{eqnarray}
where we dropped subscript $x$ for ${\bf M}_{12}$ and ${\bf M}_{21}$. 
Then the dispersion decoupling transformation (the ${\bf U}$ matrix) can be applied, followed 
by a second decoupling transformation ${\bf V}$. 
The same procedure can be carried out as for the vertical crab cavity case. 
It is noted that if the crab cavity location (point 2) is dispersion free, then the 
decoupling transfer matrix ${\bf V}$ is the same as the $y$-$z$ plane for the vertical case. 
Eqs.~(\ref{eqtC2form}-\ref{eqC22} ) are valid with the $y$-plane parameters replaced by the 
horizontal counterparts. 

\section{\label{secSimul}Numeric example and simulation}
We use the SPEAR3 storage ring lattice to work out a numeric example 
in order to illustrate the results derived in the previous sections. 
Table~\ref{tabParaSP3} lists a few related parameters of the machine. 
The method of using two crab cavities with different frequencies to tilt the beam in the 
$y$-$z$ plane for the generation of short X-ray pulse is considered~\cite{Zholents2015111}. 
For example, if the frequencies of the two crab cavities are 6 and 6.5 times of the RF frequency 
of the ring, respectively, and the deflecting voltages are properly matched, 
the tilting effects of the two crab cavities cancel for half of the buckets. 
For the other half buckets, the tilting effects add up. 
For a bunch in a tilted bucket, the linear dynamics is not different from 
the case with one crab cavity - only that the strength parameter now is the sum of the two 
crab cavities, $\epsilon=e(V_1 k_1+V_2 k_2)/E_0$, where $V_{1,2}$ and $k_{1,2}$ are the deflecting voltage and 
angular wave number for the two crab cavities, respectively. 
\begin{table}[hbt] 
\caption{ Selected Parameters of SPEAR3 }
\label{tabParaSP3}
 \begin{center}  
  \begin{tabular*}{0.4\textwidth}%
     {@{\extracolsep{\fill}}l|c|l}
  \hline
  Parameters & Value & Unit \\
  \hline  
  Energy  & $3$  & GeV    \\
  Circumference & 234.1 & m \\
  Tune $\nu_{x,y}$ & 14.106, 6.177 & \\
  RF frequency $f_{\rm rf}$ & 476.3 & MHz \\
  Bunch length $\sigma_{z}$ & 6.0 & mm \\  
  Momentum spread $\sigma_{\delta}$ & 0.001 &  \\  
  Momentum compaction $\alpha_c$ & $1.62\times10^{-3}$ & \\
  Synchrotron tune $\nu_s$ & 0.010 & \\
  \hline
  \end{tabular*}
  \end{center}
\end{table}  

In the following we consider only one crab cavity, with a deflecting voltage of $V=2$~MV 
and the frequency is $f_1=6f_{\rm rf}=2857.8$~MHz. 
The strength parameter is thus $\epsilon = -0.0399$, where there is a negative sign due to the 
choice of crab cavity phase. 
The crab cavity is located in one of the matching straight section, where 
$\beta_y=2.803$~m, $\alpha_y=-0.348$, $D_x=0.085$, and $D'_x=-0.001$. 

We consider an observation point at the center of a standard straight section (13S), where 
$\beta_y=4.860$~m, $\alpha_y=0.0$, $D_x=0.10$, $D'_x=0.0$, and the vertical betatron phase advance 
from the crab cavity to 13S is $\Delta\Psi_{y}=2.2345$ rad modulo $2\pi$. The one-turn
transfer matrix at this point is 
\begin{eqnarray}
{\bf T}_{13S} = \phantom{take space with random words, space space space space} \nonumber \\
\begin{pmatrix}
    0.7860  &  5.4926  &  0.002 &  -0.0213 &   0.001 &   0.019 \\
   -0.0699  &  0.7838  &  0.000 &  -0.0007  &  0.000  &  0.007 \\
    0.0010 &  -0.0137  &  0.439  &  4.3579  &  0.116  & -0.014\\
   -0.0002 &   0.0022 &  -0.189  &  0.4430 &  -0.019  &  0.002\\
   -0.0067 &  -0.0224 &  -0.003  &  0.0346  &  0.997  & -0.379\\
    0.0 &  -0.0008  &  0.013 &  -0.1329 &   0.010  &  0.999
\end{pmatrix}. \nonumber
\end{eqnarray}
Following the equations in section~\ref{secyz}, the matrices ${\bf C}_1$ 
and ${\bf C}_2$ are calculated as 
\begin{eqnarray}
{\bf C}_1 &=& \epsilon \begin{pmatrix}  -0.0452 &  -0.2162 \\
    0.0083 &   0.0109 \end{pmatrix},\nonumber \\
{\bf C}_2 &=& \epsilon \begin{pmatrix}   0.3799  &  1.0904 \\
    0.7146 &   0.0268\end{pmatrix}, \nonumber 
\end{eqnarray}
and the new transfer matrix is 
\begin{eqnarray}
{\bf V}^{-1}{\bf U}^{-1}{\bf T}_{\rm 13S} {\bf U}{\bf V} = \phantom{take space with random words, space } \nonumber \\
\begin{pmatrix}
    0.7860  &  5.4926  &  0.0 &  0.0  &  0.0001  &  0.0 \\
   -0.0699  &  0.7838  &  0.0  & 0.0  &  0.0 &  0.0 \\
    0.0  &  0.0  &  0.440 &   4.3526  &  0.0002  &  0.0\\
   0.0  & 0.0 &  -0.185  &  0.4439 &  -0.0002 &   0.0\\
   0.0 &   0.0 &  0.0 &   0.0002 &   0.9963 &  -0.379\\
    0.0001  &  0.0  & -0.0 &  -0.0008  &  0.0138 &   0.998
\end{pmatrix}. \nonumber
\end{eqnarray}
The off-diagonal blocks of the new transfer matrix
are substantially reduced toward zero, which verifies 
the results in section~\ref{secyz}. 

We also performed particle tracking simulation to determine the equilibrium distribution with 
the crab cavity in the lattice. 
There is no $x$-$y$ coupling in the model originally without the crab cavity. 
Simulation is done with the tracking code 
Accelerator Toolbox~\cite{AT}, with new 
functions added to model the crab cavity and quantum excitation. 
The code Elegant~\cite{Borland_elegant} is also used for tracking and good agreement is found between 
the two codes. 
All particles are launched with zero coordinate offsets and tracked for 30000 turns, which are 7.5 times of 
the longitudinal damping time and 4.4 times of the vertical damping time. 

\begin{figure}[hbt]
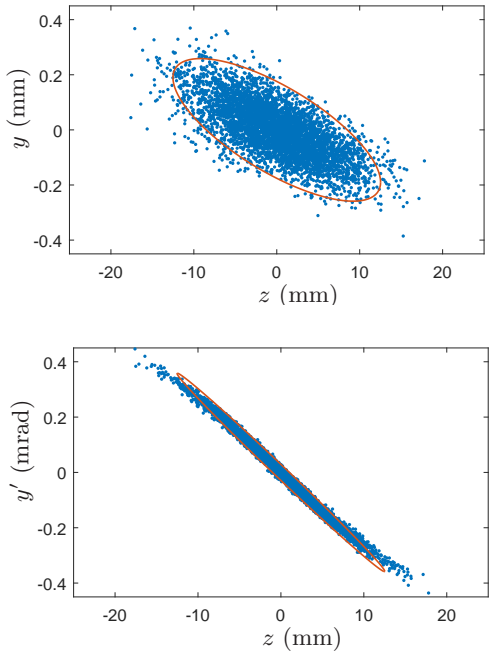

\psfrag{zmm}{\small{$z$~(mm)}}
\psfrag{ymm}{\small{$y$~(mm)}}
\psfrag{ypmrad}{\small{$y'$~(mrad)}}
\vspace*{2mm}
  \centering
  \includegraphics[width=2.5in]{fig2a.eps}
\vspace*{5mm}

  \includegraphics[width=2.5in]{fig2b.eps}

\flushleft
  \caption{\label{figYZYYp} The projection of equilibrium particle distribution (blue dots) at 13S 
  onto the $y$-$z$ plane (top) and $y'$-$z$ plane (bottom) are 
  compared to calculated ellipses (red) with Eqs.~(\ref{eqSigmayz3}-\ref{eqSigmayy3}) and 
Eqs. (\ref{eqApproxSigz},\ref{eqEigenEmitY})) (with ellipses 
  covering $6\sigma$ of the Gaussian distribution).   
  }
\end{figure}

The projection of the phase space volume onto the $y$-$z$ and $y'$-$z$ planes are 
shown in FIG.~\ref{figYZYYp} for the 13S observation point. Also plotted in the figure are the  
ellipses for the corresponding second order moments calculated with Eqs.~(\ref{eqSigmayz3}-\ref{eqSigmayy3}) and 
Eqs. (\ref{eqApproxSigz},\ref{eqEigenEmitY})). 
The area of the ellipses are 6 times of the respective projected emittances. 
Ellipses derived with numeric Ohmi envelope calculation are not shown since they overlap almost exactly with the 
ones calculated with formulas. 
There is an excellent agreement between the tracked particle distribution and the prediction 
based on the decoupling matrix for the $y$-$z$, $y'$-$z$ planes, except deviation at the tails of the distribution 
caused by the nonlinearity of the sinusoidal wave on the crab cavity. 

\begin{figure}[hbt]
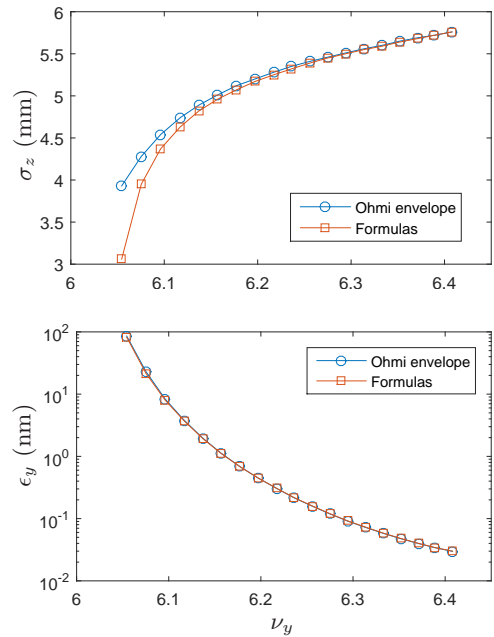

\psfrag{nuy}{\small{$\nu_y$}}
\psfrag{epsynm}{\small{$\epsilon_y$~(nm)}}
\psfrag{sigmazmm}{\small{$\sigma_z$~(mm)}}
  \centering
  \includegraphics[width=2.5in]{fig3a.eps}
  \includegraphics[width=2.5in]{fig3b.eps}
  \caption{\label{figEpsyNuy} Bunch length (top) and vertical eigen-emittance (bottom) 
  from numeric Ohmi envelope calculation is compared to formulas (Eq. (\ref{eqApproxSigz}) 
  for bunch length and Eq. (\ref{eqEigenEmitY}) for vertical eigen-emittance). 
  }
\end{figure}

To check the formulas for bunch length (Eq. (\ref{eqApproxSigz})) and vertical eigen-emittance 
(Eq. (\ref{eqEigenEmitY})), we did numeric calculation of Ohmi envelope while varying the vertical tune 
of the lattice. The results are compared to calculations by the formulas and are shown in 
FIG. \ref{figEpsyNuy}. 
It is seen that the semi-empirical formula, Eq. (\ref{eqApproxSigz}), agrees with numeric calculations 
for large tune separation between the vertical and longitudinal directions, but deviates 
from numeric results as the vertical tune approaches the synchrotron tune. 
However, the analytic formula Eq. (\ref{eqEigenEmitY}) agrees with numeric results excellently 
in the entire parameter range. 

\begin{figure}[hbt]
\psfrag{sigzmps}{\small{$\sigma_{zm}$~(ps)}}
\psfrag{flux}{\small{Flux (\%)}}
  \centering
  \vspace*{5mm}
  \includegraphics[width=2.5in]{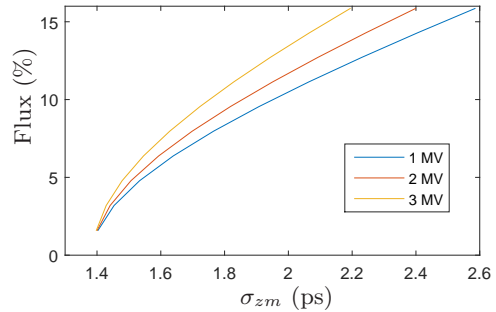}
    \caption{\label{figFluxVd} Percentage of flux vs. pulse duration for various deflecting voltages.   }
\end{figure}

The minimum pulse duration is typically dominated by the vertical eigen-emittance term in Eq. (\ref{eqMinSigt}). 
The percentage of flux for short pulses accepted by a slit as a function of pulse duration is 
plotted in FIG.~\ref{figFluxVd} for various deflecting voltages. 
No original vertical emittance or radiation divergence is assumed. In this case, indeed a higher 
deflecting voltage  does not reduce the minimum pulse duration, although it helps reduce the pulse duration 
for a given percentage of flux. 

\begin{figure}[hbt]
\psfrag{nuy}{\small{$\nu_y$}}
\psfrag{sigzmps}{\small{$\sigma_{zm}$~(ps)}}
  \centering
  \includegraphics[width=2.5in]{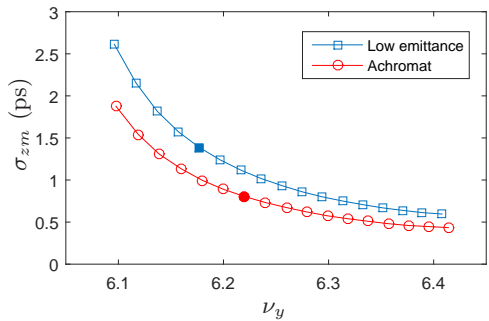}
  \caption{\label{figSigzmNuy} Minimum pulse duration vs. vertical tune for the low emittance lattice 
  and the achromat lattice of SPEAR3. Filled markers indicate nominal tunes for the lattices. 
  }
\end{figure}

Because the vertical eigen-emittance due to the crab cavity is strongly dependent on 
the momentum compaction factor (see Eqs. (\ref{eqHcAvg}-\ref{eqEigenEmitY})), 
the minimum pulse duration is expected to be sensitive to momentum compaction factor changes. 
FIG.~\ref{figSigzmNuy} compares the minimum pulse duration vs. vertical tune for two SPEAR3 lattices 
with crab cavity parameters, crab cavity and source point locations as given in the above example. 
Parameters for the ``low emittance'' lattice are listed in Table~\ref{tabParaSP3}. The ``achromat'' lattice 
has a nominal vertical tune $\nu_y=6.22$, a momentum compaction factor $\alpha_c=1.18\times10^{-3}$,  
and a nominal bunch length of $\sigma_z=5.0$~mm. 
Clearly lowering the momentum compaction factor helps reduce the pulse duration, by a factor more than the 
reduction of nominal bunch length. 
Increasing the vertical tune reduces the minimum pulse duration, although the return diminishes 
as the tune shifts up. 

\section{\label{secConclu}Conclusion}
We studied the linear coupling between the transverse and longitudinal directions 
introduced by a crab cavity in a storage ring. 
A matrix perturbation method is applied to derive the transformation that decouples 
the 6D one-turn transfer matrix. 
Analytic formulas are given for the coefficients of the decoupling transformation. 
The equilibrium particle distribution in an electron storage ring is also derived by applying the 
perturbation method to the Ohmi envelope equation~\cite{Ohmi.PRE.49.751}. 
Considering the quantum excitation in bending magnets for a beam distribution with 
tilt across the $y$-$z$ directions, we derived the vertical eigen-emittance due to crab 
cavities. 
Application to the short pulse generation scheme using crab cavities is considered. 
Numeric example and particle tracking are shown to demonstrate the analytic results.

\begin{acknowledgments}
Discussions with Sasha Zholents and James Safranek were helpful. 
The study is supported by DOE Contract No. DE-AC02-76SF00515. 
\end{acknowledgments}

\bibliography{da_ref}

\begin{thebibliography}{10}%
\makeatletter
\providecommand \@ifxundefined [1]{%
 \@ifx{#1\undefined}
}%
\providecommand \@ifnum [1]{%
 \ifnum #1\expandafter \@firstoftwo
 \else \expandafter \@secondoftwo
 \fi
}%
\providecommand \@ifx [1]{%
 \ifx #1\expandafter \@firstoftwo
 \else \expandafter \@secondoftwo
 \fi
}%
\providecommand \natexlab [1]{#1}%
\providecommand \enquote  [1]{``#1''}%
\providecommand \bibnamefont  [1]{#1}%
\providecommand \bibfnamefont [1]{#1}%
\providecommand \citenamefont [1]{#1}%
\providecommand \href@noop [0]{\@secondoftwo}%
\providecommand \href [0]{\begingroup \@sanitize@url \@href}%
\providecommand \@href[1]{\@@startlink{#1}\@@href}%
\providecommand \@@href[1]{\endgroup#1\@@endlink}%
\providecommand \@sanitize@url [0]{\catcode `\\12\catcode `\$12\catcode
  `\&12\catcode `\#12\catcode `\^12\catcode `\_12\catcode `\%12\relax}%
\providecommand \@@startlink[1]{}%
\providecommand \@@endlink[0]{}%
\providecommand \url  [0]{\begingroup\@sanitize@url \@url }%
\providecommand \@url [1]{\endgroup\@href {#1}{\urlprefix }}%
\providecommand \urlprefix  [0]{URL }%
\providecommand \Eprint [0]{\href }%
\providecommand \doibase [0]{http://dx.doi.org/}%
\providecommand \selectlanguage [0]{\@gobble}%
\providecommand \bibinfo  [0]{\@secondoftwo}%
\providecommand \bibfield  [0]{\@secondoftwo}%
\providecommand \translation [1]{[#1]}%
\providecommand \BibitemOpen [0]{}%
\providecommand \bibitemStop [0]{}%
\providecommand \bibitemNoStop [0]{.\EOS\space}%
\providecommand \EOS [0]{\spacefactor3000\relax}%
\providecommand \BibitemShut  [1]{\csname bibitem#1\endcsname}%
\let\auto@bib@innerbib\@empty
\bibitem [{\citenamefont {Zholents}(2015)}]{Zholents2015111}%
  \BibitemOpen
  \bibfield  {author} {\bibinfo {author} {\bibfnamefont {A.}~\bibnamefont
  {Zholents}},\ }\href {\doibase http://dx.doi.org/10.1016/j.nima.2015.07.016}
  {\bibfield  {journal} {\bibinfo  {journal} {Nuclear Instruments and Methods
  in Physics Research Section A: Accelerators, Spectrometers, Detectors and
  Associated Equipment}\ }\textbf {\bibinfo {volume} {798}},\ \bibinfo {pages}
  {111 } (\bibinfo {year} {2015})}\BibitemShut {NoStop}%
\bibitem [{\citenamefont {Oide}\ and\ \citenamefont
  {Yokoya}(1989)}]{Oide.PRA.40.315}%
  \BibitemOpen
  \bibfield  {author} {\bibinfo {author} {\bibfnamefont {K.}~\bibnamefont
  {Oide}}\ and\ \bibinfo {author} {\bibfnamefont {K.}~\bibnamefont {Yokoya}},\
  }\href {\doibase 10.1103/PhysRevA.40.315} {\bibfield  {journal} {\bibinfo
  {journal} {Phys. Rev. A}\ }\textbf {\bibinfo {volume} {40}},\ \bibinfo
  {pages} {315} (\bibinfo {year} {1989})}\BibitemShut {NoStop}%
\bibitem [{\citenamefont {Zholents}\ \emph {et~al.}(1999)\citenamefont
  {Zholents}, \citenamefont {Heimann}, \citenamefont {Zolotorev},\ and\
  \citenamefont {Byrd}}]{Zholents1999385}%
  \BibitemOpen
  \bibfield  {author} {\bibinfo {author} {\bibfnamefont {A.}~\bibnamefont
  {Zholents}}, \bibinfo {author} {\bibfnamefont {P.}~\bibnamefont {Heimann}},
  \bibinfo {author} {\bibfnamefont {M.}~\bibnamefont {Zolotorev}}, \ and\
  \bibinfo {author} {\bibfnamefont {J.}~\bibnamefont {Byrd}},\ }\href {\doibase
  http://dx.doi.org/10.1016/S0168-9002(98)01372-2} {\bibfield  {journal}
  {\bibinfo  {journal} {Nuclear Instruments and Methods in Physics Research
  Section A: Accelerators, Spectrometers, Detectors and Associated Equipment}\
  }\textbf {\bibinfo {volume} {425}},\ \bibinfo {pages} {385 } (\bibinfo {year}
  {1999})}\BibitemShut {NoStop}%
\bibitem [{\citenamefont {Sun}\ \emph {et~al.}(2009)\citenamefont {Sun},
  \citenamefont {Assmann}, \citenamefont {Barranco}, \citenamefont {Tom\'as},
  \citenamefont {Weiler}, \citenamefont {Zimmermann}, \citenamefont {Calaga},\
  and\ \citenamefont {Morita}}]{Yipeng.PRSTAB_2009}%
  \BibitemOpen
  \bibfield  {author} {\bibinfo {author} {\bibfnamefont {Y.-P.}\ \bibnamefont
  {Sun}}, \bibinfo {author} {\bibfnamefont {R.}~\bibnamefont {Assmann}},
  \bibinfo {author} {\bibfnamefont {J.}~\bibnamefont {Barranco}}, \bibinfo
  {author} {\bibfnamefont {R.}~\bibnamefont {Tom\'as}}, \bibinfo {author}
  {\bibfnamefont {T.}~\bibnamefont {Weiler}}, \bibinfo {author} {\bibfnamefont
  {F.}~\bibnamefont {Zimmermann}}, \bibinfo {author} {\bibfnamefont
  {R.}~\bibnamefont {Calaga}}, \ and\ \bibinfo {author} {\bibfnamefont
  {A.}~\bibnamefont {Morita}},\ }\href {\doibase 10.1103/PhysRevSTAB.12.101002}
  {\bibfield  {journal} {\bibinfo  {journal} {Phys. Rev. ST Accel. Beams}\
  }\textbf {\bibinfo {volume} {12}},\ \bibinfo {pages} {101002} (\bibinfo
  {year} {2009})}\BibitemShut {NoStop}%
\bibitem [{\citenamefont {Ohmi}\ \emph {et~al.}(1994)\citenamefont {Ohmi},
  \citenamefont {Hirata},\ and\ \citenamefont {Oide}}]{Ohmi.PRE.49.751}%
  \BibitemOpen
  \bibfield  {author} {\bibinfo {author} {\bibfnamefont {K.}~\bibnamefont
  {Ohmi}}, \bibinfo {author} {\bibfnamefont {K.}~\bibnamefont {Hirata}}, \ and\
  \bibinfo {author} {\bibfnamefont {K.}~\bibnamefont {Oide}},\ }\href {\doibase
  10.1103/PhysRevE.49.751} {\bibfield  {journal} {\bibinfo  {journal} {Phys.
  Rev. E}\ }\textbf {\bibinfo {volume} {49}},\ \bibinfo {pages} {751} (\bibinfo
  {year} {1994})}\BibitemShut {NoStop}%
\bibitem [{\citenamefont {Huang}(2007)}]{Huang.MatrixXZ.PRSTAB}%
  \BibitemOpen
  \bibfield  {author} {\bibinfo {author} {\bibfnamefont {X.}~\bibnamefont
  {Huang}},\ }\href {\doibase 10.1103/PhysRevSTAB.10.014002} {\bibfield
  {journal} {\bibinfo  {journal} {Phys. Rev. ST Accel. Beams}\ }\textbf
  {\bibinfo {volume} {10}},\ \bibinfo {pages} {014002} (\bibinfo {year}
  {2007})}\BibitemShut {NoStop}%
\bibitem [{\citenamefont {Sagan}\ and\ \citenamefont
  {Rubin}(1999)}]{SaganRubin.PRSTAB.2.074001}%
  \BibitemOpen
  \bibfield  {author} {\bibinfo {author} {\bibfnamefont {D.}~\bibnamefont
  {Sagan}}\ and\ \bibinfo {author} {\bibfnamefont {D.}~\bibnamefont {Rubin}},\
  }\href {\doibase 10.1103/PhysRevSTAB.2.074001} {\bibfield  {journal}
  {\bibinfo  {journal} {Phys. Rev. ST Accel. Beams}\ }\textbf {\bibinfo
  {volume} {2}},\ \bibinfo {pages} {074001} (\bibinfo {year}
  {1999})}\BibitemShut {NoStop}%
\bibitem [{\citenamefont {Lee}(1999)}]{Lee_AP}%
  \BibitemOpen
  \bibfield  {author} {\bibinfo {author} {\bibfnamefont {S.~Y.}\ \bibnamefont
  {Lee}},\ }\href@noop {} {\emph {\bibinfo {title} {{Accelerator physics}}}}\
  (\bibinfo  {publisher} {World Scientific},\ \bibinfo {year}
  {1999})\BibitemShut {NoStop}%
\bibitem [{\citenamefont {Terebilo}(2001)}]{AT}%
  \BibitemOpen
  \bibfield  {author} {\bibinfo {author} {\bibfnamefont {A.}~\bibnamefont
  {Terebilo}},\ }\href@noop {} {\  (\bibinfo {year} {2001})},\ \bibinfo {note}
  {{SLAC-PUB-8732}}\BibitemShut {NoStop}%
\bibitem [{\citenamefont {Borland}(2000)}]{Borland_elegant}%
  \BibitemOpen
  \bibfield  {author} {\bibinfo {author} {\bibfnamefont {M.}~\bibnamefont
  {Borland}},\ }\href@noop {} {\enquote {\bibinfo {title} {elegant: A flexible
  sdds-compliant code for accelerator simulation},}\ } (\bibinfo {year}
  {2000}),\ \bibinfo {note} {{APS Report No. LS-287}}\BibitemShut {NoStop}%
\end{thebibliography}%

\end{document}